\documentclass[aps,prb,preprint,groupedaddress,amsmath]{revtex4}
\usepackage{graphicx}

\begin{document}

\title{Response of parametrically-driven nonlinear coupled oscillators
with application to micro- and nanomechanical resonator arrays}

\author{Ron Lifshitz}
%\email[]{ronlif@post.tau.ac.il}
\affiliation{School of Physics and Astronomy, Raymond and Beverly
  Sackler Faculty of Exact Sciences, Tel Aviv University, Tel Aviv
  69978, Israel}
\author{M.~C.~Cross}
\affiliation{Condensed Matter Physics 114-36, California Institute of
  Technology, Pasadena, California 91125}
%\email[]{mcc@caltech.edu}

\date{\today}

\begin{abstract}
  The response of a coupled array of nonlinear oscillators to
  parametric excitation is calculated in the weak nonlinear limit
  using secular perturbation theory. Exact results for small arrays of
  oscillators are used to guide the analysis of the numerical
  integration of the model equations of motion for large arrays. The
  results provide a qualitative explanation for a recent experiment
  [Buks and Roukes, to appear in J.~MEMS (2002)] involving a
  parametrically-excited micromechanical resonator array. Future
  experiments are suggested that could provide quantitative tests of
  the theoretical predictions.
\end{abstract}

%\keywords{}

\maketitle



\section{\label{intro}
  Motivation: Non-linearity of MEMS and NEMS resonators}

Recent technological advances have enabled the fabrication of
mechanical resonators down to micrometer and even nanometer scales,
with frequencies almost in reach of the GHz
range.\cite{craighead,physworld}  Even though the main thrust in this
field of research comes from the need to produce smaller, lighter,
faster, and more efficient electromechanical systems, there is new
basic physics to be learned along the way.\cite{sciam} One
particularly interesting aspect of the physical behavior of micro- and
nanoelectromechanical systems (MEMS and NEMS) is their nonlinear
mechanical response at relatively small deviations from equilibrium.
This nonlinear behavior has been observed
experimentally,\cite{craighead,buks1} and also exploited to achieve
mechanical signal amplification and mechanical noise
squeezing~\cite{rugar,carr,harrington} in single resonators. In
addition, MEMS and NEMS facilitates the fabrication of large arrays of
resonators, for which the coherent response might be useful for signal
enhancement and noise reduction. It is important to understand the
nonlinear behavior of MEMS and NEMS resonators in order to improve
their future designs. At the same time, one can take advantage of
these systems for the experimental study of nonlinear dynamics.
 
This paper is motivated by a recent experiment by Buks and
Roukes~\cite[henceforth BR]{buks2} who fabricated an array of 67
fully-suspended doubly-clamped micromechanical resonating beams. Each
beam was $270\mu m\times 1\mu m\times 0.25\mu m$ in size,
and the distance between neighboring beams was $4\mu m$. The
substrate beneath the array was completely etched away, forming a
suspended diffraction grating with optical access from both sides. All
even-numbered beams were electrically connected to one electrode and
all odd-numbered beams to a second electrode. This allowed the
application of electrostatic forces to induce coupling between the
beams. The system was driven parametrically by introducing an {\it
  AC\/} component to the potential difference between the
even-numbered and odd-numbered beams. The collective response of the
array, as a function of the driving frequency and the {\it DC\/}
component of the potential difference, was measured using optical
diffraction. The response that BR inferred from their measurement was
surprising in that ({\it i\/}) instead of showing a band consisting of a
sequence of resonance peaks at the 67 normal frequencies of the array,
the typical response as the frequency was swept up showed a small
number of wide peaks where the response gradually increased and very
abruptly decreased; and ({\it ii\/}) the array responded at
frequencies beyond the expected top edge of the band. 

We show below that both of these effects are a direct result of the
fact that the restoring forces acting on the resonators as well as the
damping that they undergo are both nonlinear. In Section~\ref{eom} we
describe the simplest equations of motion that are required to model
the nonlinear resonator array. In Section~\ref{one} we solve the
response of a single nonlinear resonator to parametric excitation at
twice its resonance frequency using secular perturbation theory (for
comparison, we solve in Appendix~\ref{ap1} the response to parametric
excitation at the resonance frequency). In Section~\ref{few} we use
the same method to calculate the response of the coupled resonator
array and obtain exact results for a few (two or three) resonators.
Understanding the analytical results of these two sections allows us
to interpret the results of Section~\ref{many} where we numerically
integrate the equations of motion for an array of 67 resonators. Our
results agree qualitatively with the observations of BR, explaining
the two points mentioned above, but we suggest that further
experiments be performed in order to test our theoretical calculations
in a more quantitative fashion.

\section{\label{eom}
  Equations of motion}

We seek the simplest set of equations of motion (EOM) that capture the
important physical aspects of the array of coupled micromechanical
beams. We first note that the normal frequencies of an individual beam
are sufficiently separated that the frequency bands, formed by the
coupling of the beams in the array, are well separated by gaps in
which the system cannot respond. We therefore assume that we can treat
the lowest band separately from all the others, so that each
individual beam is oscillating strictly in its fundamental mode of
vibration. Each beam can therefore be described by a single degree of
freedom $x_n$, giving its displacement from equilibrium. We neglect
any inhomogeneities in the fabrication of the beams and assume that
all beams are identical.  BR have actually examined each beam
individually and report that their beams have a fairly uniform
distribution of resonance frequencies, with an average of
$\omega_B=179.3$kHz, and a standard deviation of 0.53kHz. There is a
much larger variation in the quality factors of the beams prior to the
application of electrostatic interaction between them, but this
variation disappears when a small potential difference is introduced
between the beams.\cite{eyal}

The coordinates $x_n$ are all assumed small so that only terms to
lowest order in $x_n$, necessary to capture the physical behavior of
the system, will be kept in the EOM.  Two types of forces act on the
beams. The elastic restoring force of each beam and the electrostatic
forces between the beams.  Experiments done by Buks and
Roukes~\cite{buks1} on single beams of the type used in the array show
that their response is like that of a Duffing oscillator---an
oscillator whose restoring force contains a term proportional to the
cube of the displacement and makes the oscillator stiffer than it
would be within the harmonic approximation.  Assuming a symmetric
restoring force, and therefore no term proportional to an even power
of $x_n$, and neglecting higher than cubic-order nonlinear
corrections, the elastic force acting on the $n^{th}$ individual beam
is
\begin{equation}
  \label{eq:elasticforce}
  F_{elastic}^{(n)} = -m\omega_{B}^{2} x_n -m\alpha x_{n}^3,
\end{equation}
where $m$ is the effective mass of a beam oscillating in its
fundamental mode, whose frequency is $\omega_B$.

Even though the electrostatic force between two parallel charged wires
decays only as $1/r$, for simplicity we consider only the attractive
interactions between nearest-neighbor beams. Within this approximation
each term in the EOM depends either on the variables $x_n$, describing
the displacement of an individual beam from its equilibrium position,
or on the difference variables $x_{n+1}-x_n$, describing the relative
displacements of a pair of neighboring beams. To keep the equations as
simple as possible we restrict each type of nonlinear term in the EOM
to depend either on $x_n$ or on $x_{n+1}-x_n$, depending on whether it
is mostly influenced by the elastic forces of the beams or the
electrostatic interaction between them, respectively.

The cubic term in the expansion of the nearest-neighbor electrostatic
interaction tends to pull the beams away from equilibrium, acting
against the cubic term in the expansion of the elastic force in
Eq.~(\ref{eq:elasticforce}). Because, as we shall confirm later, the
response of the array is consistent with having a cubic term which
stiffens the beams, the elastic contribution to the cubic term is
stronger than the electrostatic one. We therefore ignore the cubic as
well as all higher terms in the electrostatic interaction, which we
write as
\begin{equation}
  \label{eq:elecforce}
  F_{electric}^{(n)} = -\frac12 m\Delta^2[1+H\cos\omega_p t](x_{n+1} -
  2x_n +x_{n-1}). 
\end{equation}
Note that the linear electrostatic force constant $\frac12 m\Delta^2$,
which is modulated with a relative amplitude $H\ll1$, representing the
{\it DC\/} and the {\it AC\/} components of the applied voltage, is
positive, acting to soften the elastic restoring force. The factor of
$1/2$ is used with the difference variable for convenience.

Parametric excitation, as it appears in the bare Mathieu equation for
a single oscillator of frequency $\omega_0$ without all the additional
terms that we have here, is an instability of the system that occurs
whenever the drive frequency is around one of the special values
$\omega_p=2\omega_0/n$, where $n$ is an integer that labels the
so-called instability tongues of the system (named after the
tongue-shaped instability curves in the frequency-amplitude
plane).\cite{landau} We choose the parametric driving frequency
$\omega_p$ to be around twice some value $\omega_0$
within the array's band of normal frequencies. We are therefore
exciting the system in its first ($n=1$) instability tongue. Thus,
\begin{equation}
  \label{eq:omegap}
  \omega_p = 2\omega_0 + \epsilon\Omega,
\end{equation}
where $\epsilon$ is a small parameter. In the BR experiment the system
was actually excited in its second instability tongue, {\it i.e.}
$\omega_p$ was chosen around some frequency in the band. It turns out
that the response at the second tongue, apart from a few differences,
is quite similar to that of the first tongue. We therefore prefer to
carry out full calculations only for the first tongue which is
somewhat easier to handle, and just for comparison, we calculate in
Appendix~\ref{ap1} the response of a single nonlinear oscillator,
excited at its second tongue.

There is good reason to believe that most of the dissipation in the
coupled system is a result of the electrostatic interaction which
causes currents to flow through the beams. This assumption is based on
the observation of Buks and Roukes~\cite{eyal} that the quality factors
greatly diminish as the {\it DC\/} component of the electrostatic
potential is increased. We therefore make the simplifying
approximation that dissipation occurs predominantly as a result of
currents, all other dissipation mechanisms being relatively
negligible. This approximation relieves us of the worry about the
variation in the quality factors of the individual beams before
application of the electrostatic potential. The dissipative forces in
the EOM are therefore written with respect to the difference variable,
\begin{eqnarray}
  \label{eq:diss}\nonumber
  &F_{diss}^{(n)} 
   &= \frac12 m\omega_B\Gamma (\dot x_{n+1} - 2\dot x_n +\dot x_{n-1})\\
  &&+\ \frac12 m\omega_B\alpha\eta [(x_{n+1}-x_n)^2(\dot x_{n+1}-\dot x_n) -
            (x_{n}-x_{n-1})^2(\dot x_{n}-\dot x_{n-1})],
\end{eqnarray}
where we have included a nonlinear contribution to the dissipation, of
the same order as the nonlinear elastic force~(\ref{eq:elasticforce}).
When putting all the pieces together, we (a) divide out the effective
mass $m$ of a beam; (b) scale time $t\to t/\omega_B$ so that all
frequencies (including $\Delta$) are measured in units of $\omega_B$;
and (c) scale length $x\to x/\sqrt\alpha$ to get rid of the dependence
on $\alpha$. The equation of motion for the $n^{th}$ beam becomes
\begin{eqnarray} \label{eq:preeom}\nonumber
\ddot x_n &+ &x_n + x_n^3 + \frac12 \Delta^2\bigl[1 + H\cos(2\omega_0 +
    \epsilon\Omega) t\bigr](x_{n+1} - 2x_n + x_{n-1})\\ \nonumber 
&- &\frac12 \Gamma (\dot x_{n+1} - 2\dot x_n + \dot x_{n-1})\\
&- &\frac12 \eta \bigl[(x_{n+1}-x_n)^2(\dot x_{n+1} - \dot x_n)
- (x_n-x_{n-1})^2(\dot x_n - \dot x_{n-1})\bigr] = 0.
\end{eqnarray}

In the following sections we shall solve these equations using secular
perturbation theory. The physical parameter that allows us to use this
approach is the linear damping coefficient which is assumed to be
small. We therefore express it as $\Gamma=\epsilon\gamma$, taking
$\epsilon$ to be our small expansion parameter. The parametric
instability of the system then occurs for small driving amplitude near
resonance, and if, in addition, we consider the system near the onset
of the instability, we can assume that the effects of nonlinearity are
small as well. Thus, for small displacements $x_n$ all the nontrivial
physical effects, namely, the parametric excitation, the cubic elastic
restoring force, and both the linear and the amplitude-dependent
dissipation, all enter the EOM as perturbative corrections to the
simple linear equations. All these perturbative terms can be chosen to
enter the EOM in the same order of the small parameter $\epsilon$ by
taking the leading order in $x_n$ to be of order $\epsilon^{1/2}$, and
expressing $\Delta^2 H=\epsilon h$. This ensures, as we shall
confirm later on, that all the terms will contribute to the lowest-order
solution we are seeking. The final form of the EOM is therefore
\begin{eqnarray} \label{eq:eom}\nonumber
\ddot x_n &+ &x_n + x_n^3 + \frac12\bigl[\Delta^2 + \epsilon
         h\cos(2\omega_0 + \epsilon\Omega) t\bigr]  
         (x_{n+1} - 2x_n + x_{n-1})\\ \nonumber
&- &\frac12 \epsilon\gamma 
     (\dot x_{n+1} - 2\dot x_n + \dot x_{n-1})\\
&- &\frac12 \eta \bigl[(x_{n+1}-x_n)^2(\dot x_{n+1} - \dot x_n)
- (x_n-x_{n-1})^2(\dot x_n - \dot x_{n-1})\bigr] = 0.
\end{eqnarray}
As for boundary conditions, we follow the experiment of BR who had two
additional fixed beams, identical to all the rest, at both ends of the
array. This means that we define two extra variables and set them to
zero, $x_0=x_{N+1}=0$.

\section{\label{one}
  Response of a single parametrically-driven nonlinear oscillator}

We begin by calculating the response of a single nonlinear oscillator
to parametric excitation. Previous calculations of this problem exist
in the literature\cite[and references therein]{litak}, nevertheless,
we solve it here as a precursor to the many-oscillator case, treated
in the next section.  The equation of motion~(\ref{eq:eom}) for the
single-oscillator case becomes
\begin{eqnarray}
  \label{eq:single}\nonumber
  \ddot x &+ &\bigl[\omega^2 - \epsilon h\cos(2\omega +
  \epsilon\Omega) t\bigr] x \\
  &+ &\epsilon\gamma \dot x + x^3 + \eta x^2 \dot x = 0,
\end{eqnarray}
where we choose $\omega_0$ to be $\omega=\sqrt{1-\Delta^2}$, the
resonance frequency of the beam in the harmonic approximation. The
parametric excitation is performed around twice the actual resonance
frequency of the oscillator. (In Appendix~\ref{ap1} we treat the case
where the excitation is performed around the resonance frequency of
the resonator.) 

We calculate the correction to linear response by using secular
perturbation theory.\cite{hfsecular,strogatz} Recalling that the
motion of the oscillator away from equilibrium is on the order of
$\epsilon^{1/2}$ we try a solution of the form
\begin{equation}
  \label{eq:ansatz1}
  x(t) = \epsilon^{1/2}(A(T) e^{i\omega t} + c.c.) 
       + \epsilon^{3/2}x_1(t) + \ldots
\end{equation}
where $T=\epsilon t$ is a slow time variable, allowing the complex
amplitude $A(T)$ to vary slowly in time. The variation of $A(T)$ gives
us the extra freedom to eliminate secular terms and ensure that the
perturbative correction $x_1(t)$, as well as all higher-order
corrections to the linear response, do not diverge (as they do if one
uses naive perturbation theory). Using the relation
\begin{equation}
  \label{eq:adot}
  \dot A = {dA\over dt} = \epsilon {dA\over dT} \equiv \epsilon A',
\end{equation}
we calculate the time derivatives of the trial
solution~(\ref{eq:ansatz1})
\begin{subequations}\label{eq:derivs}
\begin{eqnarray}
  \dot x &= &\epsilon^{1/2}\left([i\omega A + \epsilon A']e^{i\omega t}
       + c.c.\right)  
       + \epsilon^{3/2}\dot x_1(t) + \ldots\\
  \ddot x &= &\epsilon^{1/2}\left([-\omega^2 A + 2 i\omega\epsilon A' 
        + \epsilon^2 A'']e^{i\omega t} + c.c.\right)  
       + \epsilon^{3/2}\ddot x_1(t) + \ldots
\end{eqnarray}
\end{subequations}
Substituting these expressions back into the equation of
motion~(\ref{eq:single}), and picking out all terms of order
$\epsilon^{3/2}$, we get the following equation for the first
perturbative correction
\begin{eqnarray}
  \label{eq:xone} \nonumber
  \ddot x_1 &+ &\omega^2 x_1 = -\left(2 i\omega A'e^{i\omega t} +
    c.c.\right) 
    + h\cos[(2\omega + \epsilon\Omega) t] \left(A e^{i\omega t} +
    c.c.\right)\\
    &- &\gamma \left(i\omega A e^{i\omega t} + c.c.\right)
    - \left(A e^{i\omega t} + c.c.\right)^3
    - \eta\left(A e^{i\omega t} + c.c.\right)^2  \left(i\omega A e^{i\omega t}
    + c.c.\right).
\end{eqnarray}

The collection of terms proportional to $e^{i\omega t}$ on the
right-hand side of Eq.~(\ref{eq:xone}), called the secular terms, act
like a force, driving the simple harmonic oscillator on the left-hand
side at its resonance frequency. The sum of all the secular terms must
vanish so that the perturbative correction $x_1(t)$ will not diverge.
This gives us an equation for determining the slowly varying amplitude
$A(T)$. After expressing the cosine as a sum of exponentials we get
\begin{equation}
  \label{eq:secular1}
  2i\omega {dA\over dT} - {h\over2} A^* e^{i\Omega T}
  + i\omega\gamma A + 3|A|^2 A + i\omega\eta |A|^2 A = 0.
\end{equation}
Ignoring initial transients, and assuming that the nonlinear terms in
the equation are sufficient to saturate the growth of the instability,
we try a steady-state solution of the form
\begin{equation}
  \label{eq:A1}
  A(T) = a e^{i{\Omega\over 2} T}.
\end{equation}
The solution to the equation of motion~(\ref{eq:single}) is therefore
\begin{equation}
  \label{eq:sol1}
  x(t) = \epsilon^{1/2}(a e^{i(\omega+\epsilon\Omega/2) t} + c.c.) 
       + O(\epsilon^{3/2}),
\end{equation}
where we are not interested in the correction $x_1(t)$ of
order $\epsilon^{3/2}$, but rather in the fixed amplitude $a$ of the
lowest order term. This amplitude $a$ can be any solution of the equation
\begin{equation}
  \label{eq:amp1}
  \left[(3|a|^2 - \omega\Omega) + i\omega(\gamma + \eta
  |a|^2)\right]a = {h\over2} a^*,
\end{equation}
obtained by substituting the steady-state solution~(\ref{eq:A1}) into
the equation~(\ref{eq:secular1}) of the secular terms. We immediately
see that having no response ($a=0$) is always a possible solution
regardless of the excitation frequency $\Omega$. We divide both
sides of the last equation by $\gamma\omega$ and define the rescaled
variables $\bar a=a/\sqrt{\gamma\omega}$, $\bar\Omega=\Omega/\gamma$,
$\bar\eta=\omega\eta$, and $\bar h=h/2\gamma\omega$, in terms of which
the equation for the fixed complex amplitude $a$ becomes
\begin{equation}
  \label{eq:scaledamp1}
  \left[(3|\bar a|^2 - \bar\Omega) + i(1 + \bar\eta
  |\bar a|^2)\right]\bar a = \bar h \bar a^*.
\end{equation}
Expressing $\bar a=|\bar a|e^{i\phi}$ we obtain, after taking the magnitude
squared of both sides, the intensity $|\bar a|^2$ of the non-trivial
response as all positive roots of the equation
\begin{equation}
  \label{eq:response1}
  \left({3|\bar a|^2} - \bar\Omega\right)^2 
  + \left(1 + \bar\eta|\bar a|^2\right)^2 = \bar h^2.
\end{equation}
This has the form of a distorted ellipse in the $(\bar\Omega,|\bar
a|^2)$ plane, and a parabola in the $(|\bar a|^2,\bar h)$ plane. In
addition, we obtain for the relative phase of the response 
\begin{equation}
  \label{eq:phase1}
  \phi={i\over2}\ln{{\bar a^*}\over{\bar a}}
  = -{1\over2}\arctan{{1+\bar\eta|\bar a|^2}\over{3|\bar a|^2 -
  \bar\Omega}}. 
\end{equation}

\begin{figure}
\includegraphics{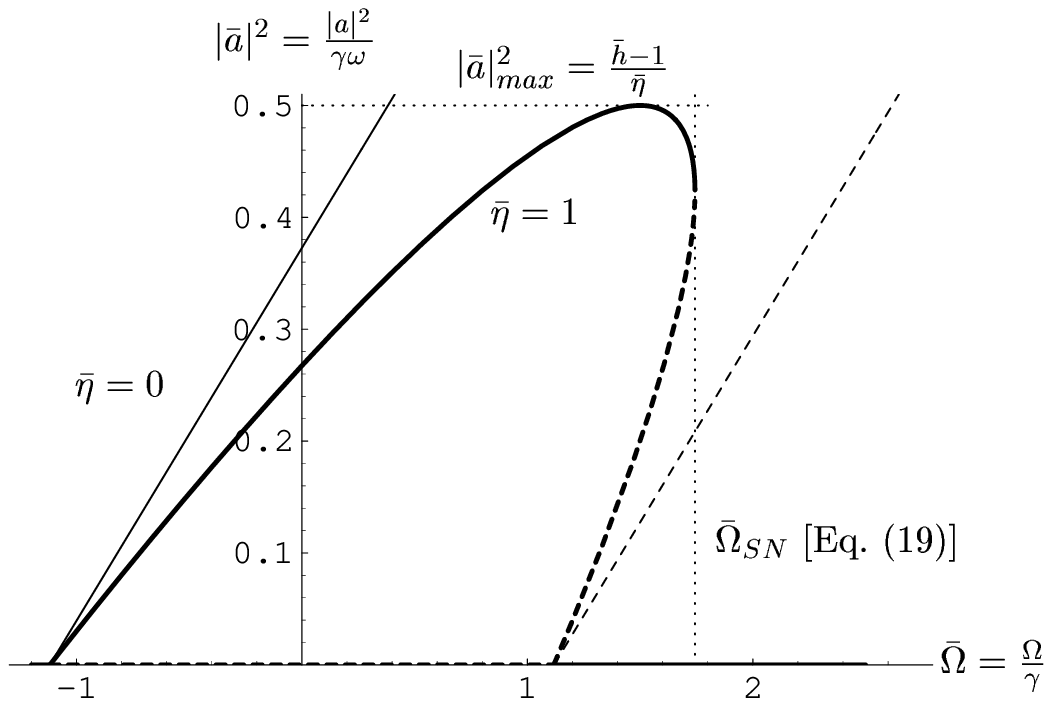}%
\caption{\label{resp1omega}
  Response intensity $|\bar a|^2$ as a function of the frequency
  $\bar\Omega$, for fixed amplitude $\bar h=1.5$.  Solid curves are
  stable solutions; dashed curves are unstable solutions. Thin curves
  show the response without non-linear damping ($\bar\eta=0$). Thick
  curves show the response for finite nonlinear damping
  ($\bar\eta=1$). Dotted lines indicate the maximal response intensity
  $|\bar a|^2_{max}$ and the saddle-node frequency $\bar\Omega_{SN}$.}
\end{figure}

\begin{figure}
\includegraphics{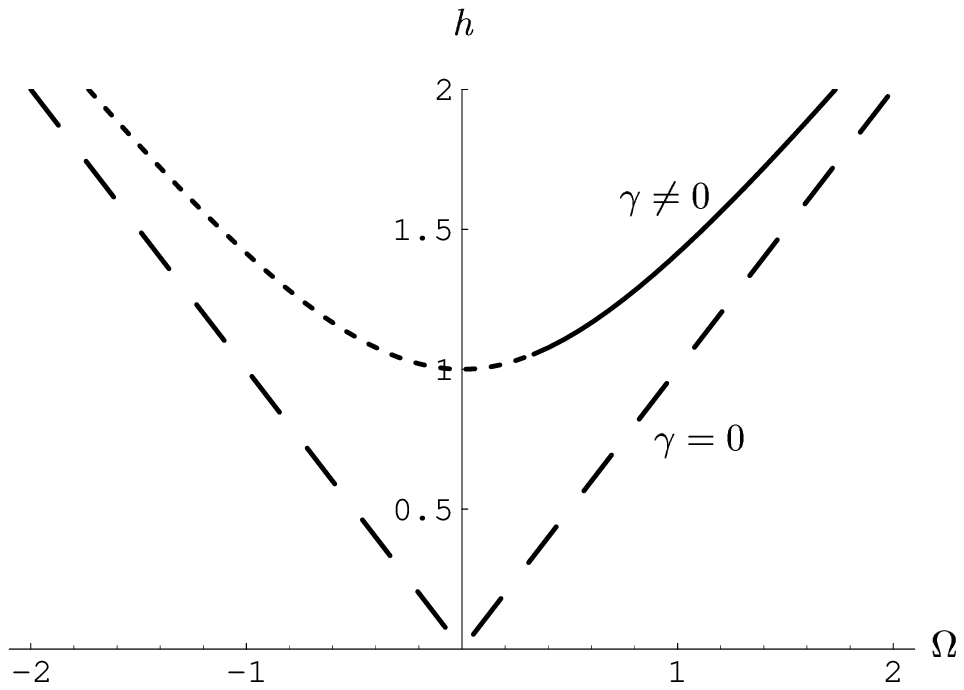}%
\caption{\label{inst1}
  Threshold for instability plotted in the $(\Omega,h)$ plane.  The
  lower, long-dashed curve shows the threshold without any linear
  damping ($\gamma=0$), which is zero on resonance. The upper curve
  shows the threshold with linear damping ($\gamma\neq0$). The
  parameters for the upper curve are $\omega=1/2$ and $\gamma=1$ so
  that $\bar h=h$. The threshold on resonance ($\bar\Omega=\Omega=0$)
  is therefore $\bar h=h=1$. The solid and short-dashed regions of the
  upper curve indicate the so-called subcritical and supercritical
  branches of the instability, respectively. On the subcritical branch
  ($\bar\Omega>\bar\eta/3$) there will be hysteresis as $h$ is varied,
  and on the supercritical branch ($\bar\Omega<\bar\eta/3$) there will
  not be any hysteresis. }
\end{figure}

\begin{figure}
\includegraphics{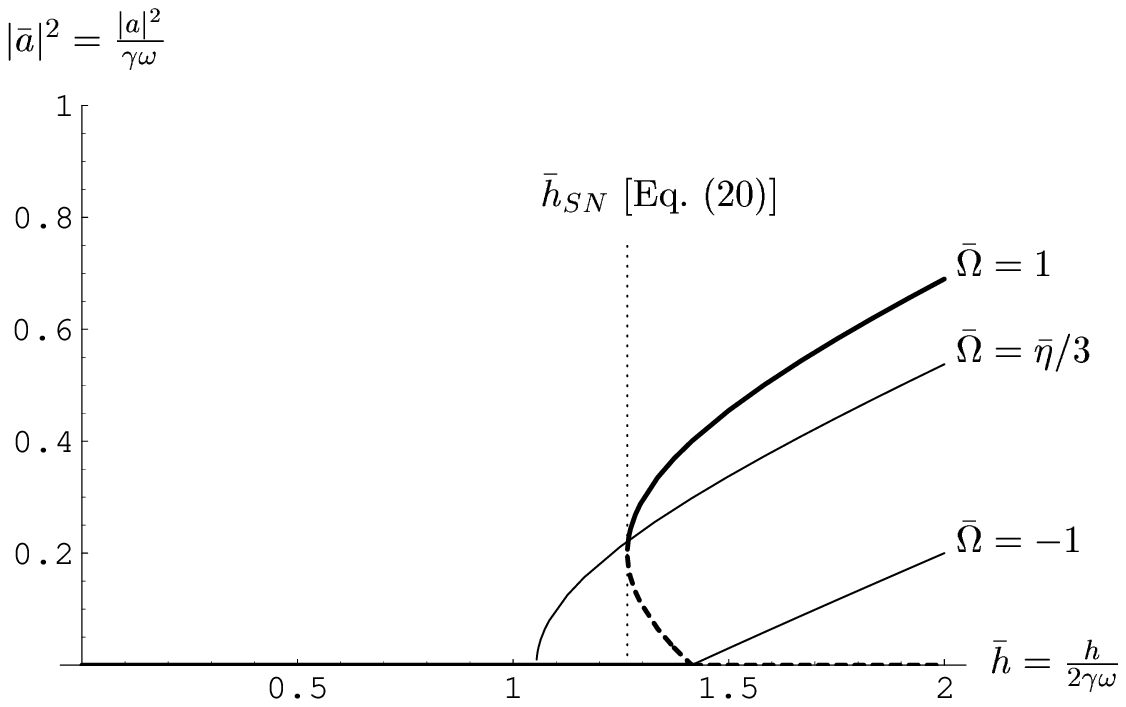}%
\caption{\label{resp1h}
  Response intensity $|\bar a|^2$ as a function of the parametric
  modulation amplitude $\bar h$ for fixed frequency $\bar\Omega$ and
  finite nonlinear damping ($\bar\eta=1)$.  Thick curves show the
  stable (solid curves) and unstable (dashed curves) response for
  $\bar\Omega=1$. Thin curves show the stable solutions for
  $\bar\Omega=\bar\eta/3$ and $\bar\Omega=-1$, and demonstrate that
  hysteresis as $\bar h$ is varied is expected only for $\bar\Omega >
  \bar\eta/3$.  }
\end{figure}

In Figure~\ref{resp1omega} we plot the response intensity $|\bar a|^2$
of a single oscillator to parametric excitation as a function of
frequency $\bar\Omega$, for fixed amplitude $\bar h=1.5$, in terms of
the rescaled variables. Solid curves indicate stable solutions, and
dashed curves are solutions that are unstable to small perturbations.
Thin curves show the response without nonlinear damping ($\bar\eta=0$)
which grows indefinitely with frequency $\bar\Omega$ and is therefore
incompatible with the experimental observations of BR and the
assumptions of our calculation.  Thick curves show the response with
finite nonlinear damping ($\bar\eta=1$).  With finite $\bar\eta$ there
is a maximum value for the response $|\bar a|^2_{max}=(\bar
h-1)/\bar\eta$, and a maximum frequency,
\begin{equation}
  \label{eq:omegafalloff}
  \bar\Omega_{SN} = \bar h
  \sqrt{1+\left(3\over{\bar\eta}\right)^2} - {3\over{\bar\eta}},
\end{equation}
at which the stability of the solution changes (known as a saddle-node
bifurcation). For frequencies above $\bar\Omega_{SN}$ the only
solution is the trivial one $\bar a=0$. These values are indicated by
horizontal and vertical dotted lines in Figure~\ref{resp1omega}.

The threshold for the instability of the trivial solution is easily
calculated by setting $\bar a=0$ in the expression
(\ref{eq:response1}) for the nontrivial solution. As seen in
Figure~\ref{resp1omega}, for a given $\bar h$ the threshold is
situated at $\bar\Omega=\pm\sqrt{{\bar h}^2 - 1}$.  The threshold is
plotted in Figure~\ref{inst1}, in the $(\Omega,h)$ plane. Note that
the minimal amplitude needed for instability is obtained on resonance
($\bar\Omega=0$) and its value is $\bar h=1$, or $h=2\gamma\omega$, so
that it scales as the linear damping coefficient $\gamma$.

Finally, in Figure~\ref{resp1h} we plot the response intensity $|\bar
a|^2$ of the oscillator as a function of amplitude $\bar h$, for fixed
frequency $\bar\Omega$ and finite nonlinear damping $\bar\eta=1$.
Again, solid curves indicate stable solutions, and dashed curves
unstable solutions. Thick curves show the response for $\bar\Omega=1$,
and thin curves show the response for $\bar\Omega=\bar\eta/3$ and
$\bar\Omega=-1$. The intersection of the trivial and the nontrivial
solutions, which corresponds to the instability threshold, occurs at
$\bar h=\sqrt{{\bar\Omega}^2 + 1}$.  For $\bar\Omega<\bar\eta/3$ the
nontrivial solution for $|\bar a|^2$ grows continuously for $\bar h$
above threshold and is stable. This is a supercritical bifurcation. On
the other hand, for $\bar\Omega > \bar\eta/3$ the bifurcation is
subcritical---the nontrivial solution grows for $\bar h$ {\it below\/}
threshold. This solution is unstable until the curve of $|\bar a|^2$
as a function of $\bar h$ bends around at a saddle-node bifurcation at
\begin{equation}
  \label{eq:hfalloff}
  \bar h_{SN} = {{1 + {\bar\eta\over3}\bar\Omega}\over\sqrt{1 +
  \left({\bar\eta\over3}\right)^2}},  
\end{equation}
where the solution becomes stable and $|\bar a|^2$ is once more an
increasing function of $\bar h$. For amplitudes $\bar h < \bar h_{SN}$
the only solution is the trivial one $\bar a=0$.

Like the response of a forced Duffing oscillator, the response of a
parametrically excited Duffing oscillator also exhibits hysteresis in
a frequency scan. If the frequency $\bar\Omega$ starts out at negative
values and is increased gradually with a fixed amplitude $\bar h$, the
response will gradually increase along the thick solid curve in
Figure~\ref{resp1omega}, until $\bar\Omega$ reaches $\bar\Omega_{SN}$
and the response drops abruptly to zero. If the frequency is then
decreased gradually, the response will remain zero until $\bar\Omega$
reaches the upper instability threshold $\sqrt{\bar h^2-1}$, and the
response will jump abruptly to the thick solid curve above, and then
gradually decrease to zero along this curve. A similar hysteretic
behavior will be observed if the amplitude $\bar h$ is varied with a
fixed frequency $\bar\Omega>\bar\eta/3$, as can be inferred from
Figure~\ref{resp1h}.

\section{\label{few}
  Response of a parametrically-driven array of nonlinear coupled 
  oscillators---Secular perturbation theory}

Consider now the coupled array of nonlinear oscillators as described
by the general EOM~(\ref{eq:eom}). We calculate its response to
parametric excitation, again using secular perturbation theory. We
expand $x_n(t)$ as a sum of standing wave modes with slowly varying
amplitudes
\begin{equation}
  \label{eq:ansatz}
  x_n(t) = \epsilon^{1/2}\sum_{m=1}^N 
         \left(A_m(T) \sin(nq_m)e^{i\omega_m t} + c.c.\right)  
       + \epsilon^{3/2}x_n^{(1)}(t) + \ldots,\quad n=1\ldots N.
\end{equation}
Recall that the boundary conditions are such that there are two
additional fixed beams, labeled 0 and $N+1$, exerting electrostatic
forces on the first and the last beams of the array.  With these
boundary conditions ($x_0=x_{N+1}=0$), the possible wave vectors $q_m$
are given by
\begin{equation}
  \label{eq:bc}
  q_m = {{m\pi}\over{N+1}},\qquad m=1\ldots N.
\end{equation}

We substitute the trial solution~(\ref{eq:ansatz}) into the
EOM term by term. Up to order $\epsilon^{3/2}$ we have:
\begin{subequations}
\begin{equation}
  \label{eq:xnddot}
    \ddot x_n = \epsilon^{1/2}\sum_m\sin(nq_m)\left([-\omega^2 A_m + 2
       i\omega\epsilon A'_m]e^{i\omega_m t} + c.c.\right)   
       + \epsilon^{3/2}\ddot x_n^{(1)}(t),
\end{equation}
\begin{eqnarray}
  \label{eq:diff}\nonumber
  x_{n+1}-2x_n+x_{n-1} &=&-4 \epsilon^{1/2}\sum_m 
  \sin^2\left(q_m\over 2\right) \sin(nq_m)\left(A_m e^{i\omega_mt}+
  c.c.\right)\\ 
  &+&\epsilon^{3/2}\left(x^{(1)}_{n+1}-2x^{(1)}_n+
  x^{(1)}_{n-1}\right),
\end{eqnarray}
\begin{equation}
  \label{eq:dotdiff}
  \frac12 \epsilon\gamma 
     (\dot x_{n+1} - 2\dot x_n + \dot x_{n-1}) = -2\epsilon^{3/2}\gamma
  \sum_m \omega_m \sin^2\left(q_m\over 2\right) \sin(nq_m)\left(iA_m
     e^{i\omega_mt} + c.c.\right),
\end{equation}
\begin{eqnarray}
  \label{eq:cube}\nonumber
  x_n^3 &= &\epsilon^{3/2}\sum_{j,k,l} \sin(nq_j) \sin(nq_k)
    \sin(nq_l)\\ \nonumber 
    &\times&\left(A_je^{i\omega_jt} + c.c.\right)
    \left(A_ke^{i\omega_kt} + c.c.\right)
    \left(A_le^{i\omega_lt} + c.c.\right)\\ \nonumber
    &= &{\epsilon^{3/2}\over 4}\sum_{j,k,l}
    \left\{\sin[n(-q_j+q_k+q_l)] + \sin[n(q_j-q_k+q_l)]\right.\\ \nonumber
    &&\left.+\sin[n(q_j+q_k-q_l)] - \sin[n(q_j+q_k+q_l)]\right\}\\ 
    &\times&\left\{A_jA_kA_l e^{i(\omega_j+\omega_k+\omega_l)t} +
    3A_jA_kA_l^* e^{i(\omega_j+\omega_k-\omega_l)t} + c.c.\right\},
\end{eqnarray}
and
\begin{eqnarray}
  \label{eq:dotcube}\nonumber
  \frac12 \eta \left[(x_{n+1}-x_n)^2(\dot x_{n+1} - \dot x_n)
  \right. &-& \left.(x_n-x_{n-1})^2(\dot x_n - \dot x_{n-1})\right] \\\nonumber
  &=&-2\eta\epsilon^{3/2}\sum_{j,k,l}\sin{q_j\over2} \sin{q_k\over2}
    \sin{q_l\over2}\\ \nonumber 
  &\times&\left\{\sin\left[{{-q_j+q_k+q_l}\over2}\right]\sin[n(-q_j+q_k+q_l)]\right.\\ \nonumber
  &&+\sin\left[[{{q_j-q_k+q_l}\over2}\right]\sin[n(q_j-q_k+q_l)]\\ \nonumber
  &&+\sin\left[[{{q_j+q_k-q_l}\over2}\right]\sin[n(q_j+q_k-q_l)]\\ \nonumber
  &&\left.+\sin\left[[{{q_j+q_k+q_l}\over2}\right]\sin[n(q_j+q_k+q_l)]\right\}\\ 
  &\times&\left(A_je^{i\omega_jt} + c.c.\right)
    \left(A_ke^{i\omega_kt} + c.c.\right)
    \left(i\omega_lA_le^{i\omega_lt} + c.c.\right).
\end{eqnarray}
\end{subequations}

At order $\epsilon^{1/2}$ we simply get the linear dispersion relation,
given by
\begin{equation}
  \label{eq:dispersion}
  \omega_m^2 = 1 - 2\Delta^2 \sin^2({q_m\over 2}),\quad m=1\ldots N.
\end{equation}

At order $\epsilon^{3/2}$ we get $N$ equations of the form
\begin{equation}
  \label{eq:order3/2}
  \ddot x_n^{(1)} + x_n^{(1)} + {1\over2}\Delta^2\left(x_{n+1}^{(1)}
  -2x_n^{(1)} + x_{n-1}^{(1)}\right) = \sum_m \left(m^{th}\ {\rm
    secular\ term}\right)e^{i\omega_mt} + {\rm other\ terms},
\end{equation}
where the left-hand sides are, again, coupled linear harmonic
oscillators, with a dispersion relation given by
(\ref{eq:dispersion}).  On the right-hand sides we have $N$ secular
terms which act to drive the coupled oscillators $x_n^{(1)}$ at their
resonance frequencies.  As we did for a single oscillator in
section~\ref{one}, here too we require that all the secular terms
vanish so that the $x_n^{(1)}$ remain finite, and thus obtain equations
for the slowly varying amplitudes $A_m(T)$. To extract the equation
for the $m^{th}$ amplitude $A_m(T)$ we make use of the orthogonality
of the modes, multiplying all the terms by $\sin(nq_m)$ and summing
over $n$. We also express all normal frequencies relative to the same
reference frequency $\omega_0$, used to define the excitation
frequency $\omega_p$ in Eq.~(\ref{eq:omegap}), so that 
\begin{eqnarray}
  \label{eq:omega0}
  \omega_m = \omega_0 + \epsilon\Omega_m.
\end{eqnarray}
We find that the coefficient of the $m^{th}$ secular term, which is
required to vanish, is given by
\begin{eqnarray}
  \label{eq:Am1}\nonumber
  &-&2i\omega_m{dA_m\over{dT}} - 2i\gamma\omega_m
  \sin^2\left({q_m\over2}\right) + h A_m^*
  \sin^2\left({q_m\over2}\right) e^{i\left(\Omega -
  2\Omega_m\right)T}\\ \nonumber 
  &-&{3\over4}\sum_{j,k,l}A_jA_kA_l^* e^{i\left(\Omega_j +\Omega_k -
  \Omega_l - \Omega_m\right)T}\Delta^{(1)}_{jkl;m}\\ \nonumber
  &-&2\eta\sum_{j,k,l}\left\{\left[2i\omega_lA_j^*A_kA_l e^{i\left(-\Omega_j
  +\Omega_k + \Omega_l - \Omega_m\right)T}-i\omega_lA_jA_kA_l^*
  e^{i\left(\Omega_j +\Omega_k - \Omega_l - \Omega_m\right)T}\right]\right.\\
  &&\left.\times \Delta^{(2)}_{jkl;m} \sin{q_j\over2} \sin{q_k\over2}
    \sin{q_l\over2} \sin{q_m\over2}\right\}=0,
\end{eqnarray}
where we have introduced two $\Delta$ functions, defined in terms
of Kronecker deltas as:
\begin{subequations}
\begin{eqnarray}
  \label{eq:Delta1}\nonumber
  \Delta^{(1)}_{jkl;m}&=& \delta_{-j+k+l,m} -  \delta_{-j+k+l,-m} -
  \delta_{-j+k+l,2(N+1)-m}\\\nonumber
  &+&\delta_{j-k+l,m} -  \delta_{j-k+l,-m} -
  \delta_{j-k+l,2(N+1)-m}\\\nonumber
  &+&\delta_{j+k-l,m} -  \delta_{j+k-l,-m} -
  \delta_{j+k-l,2(N+1)-m}\\
  &-&\delta_{j+k+l,m} + \delta_{j+k+l,2(N+1)-m} - \delta_{j+k+l,2(N+1)+m},
\end{eqnarray}
and
\begin{eqnarray}
  \label{eq:Delta2}\nonumber
  \Delta^{(2)}_{jkl;m}&=& \delta_{-j+k+l,m} +  \delta_{-j+k+l,-m} -
  \delta_{-j+k+l,2(N+1)-m}\\\nonumber
  &+&\delta_{j-k+l,m} +  \delta_{j-k+l,-m} -
  \delta_{j-k+l,2(N+1)-m}\\\nonumber
  &+&\delta_{j+k-l,m} +  \delta_{j+k-l,-m} -
  \delta_{j+k-l,2(N+1)-m}\\
  &+&\delta_{j+k+l,m} -  \delta_{j+k+l,2(N+1)-m} - \delta_{j+k+l,2(N+1)+m}.
\end{eqnarray}
\end{subequations}
These $\Delta$ functions ensure the conservation of lattice
momentum---the conservation of momentum to within the non-uniqueness
of the specification of the normal modes due to the fact that
$\sin(nq_m)= \sin(nq_{2k(N+1)\pm m})$ for any integer $k$. The first
Kronecker delta in each line is a condition of direct momentum
conservation, and the other two are the so-called umklapp conditions
where only lattice momentum is conserved.

As for the single oscillator, we again try a steady-state solution, this
time of the form
\begin{equation}
  \label{eq:Am}
  A_m(T)=a_me^{i\left({\Omega\over2}-\Omega_m\right)T},
\end{equation}
so that the solutions to the EOM, after substitution of (\ref{eq:Am})
into (\ref{eq:ansatz}), become
\begin{equation}
  \label{eq:solj}
  x_n(t) = \epsilon^{1/2}\sum_m 
         \left(a_m \sin(nq_m)
         e^{i\left(\omega_0+{\epsilon\Omega\over2}\right) t} + 
         c.c.\right) + O(\epsilon^{3/2}),
\end{equation}
where all modes are oscillating at half the parametric excitation
frequency $\omega_p/2$.

As before, we are not interested in the corrections of order
$\epsilon^{3/2}$ but only in the values of the fixed amplitudes $a_m$
as functions of all the parameters of the original EOM.  Substituting
the steady state solution (\ref{eq:Am}) into the equations
(\ref{eq:Am1}) for the time-varying amplitudes $A_m(T)$, we obtain the
required equations for the fixed complex amplitudes $a_m$
\begin{eqnarray} 
  \label{eq:ampeq}\nonumber
  (\Omega-2\Omega_m)\omega_ma_m -
  2i\gamma\omega_ma_m\sin^2\left({q_m\over2}\right) +
  ha_m^*\sin^2\left({q_m\over2}\right) -
  {3\over4}\sum_{j,k,l}a_ja_ka_l^*\Delta^{(1)}_{jkl;m}\\ 
  -2i\eta\sin{q_m\over2} \sum_{j,k,l}\omega_l\left[2a_j^*a_ka_l -
  a_ja_ka_l^*\right]\sin{q_j\over2} \sin{q_k\over2}
  \sin{q_l\over2} \Delta^{(2)}_{jkl;m} =0.   
\end{eqnarray}

We can change to rescaled variables as we did in the case of a single
oscillator by dividing the equations for the amplitudes
(\ref{eq:ampeq}) by $(\gamma\omega_0)^{3/2}$ and defining as before
$\bar a_j=a_j/\sqrt{\gamma\omega_0}$, $\bar\Omega=\Omega/\gamma$,
$\bar\eta=\omega_0\eta$, and $\bar h=h/2\gamma\omega_0$, and in
addition $r_m=\omega_m/\omega_0$ and $\delta_m=2\Omega_m/\gamma$.
After doing so we obtain the rescaled equations
\begin{eqnarray} 
  \label{eq:ampscaled}\nonumber
  (\bar\Omega-\delta_m) r_m \bar a_m -
  2i r_m \sin^2\left({q_m\over2}\right) \bar a_m +
  2\bar h\sin^2\left({q_m\over2}\right) \bar a_m^* -
  {3\over4}\sum_{j,k,l}\bar a_j\bar a_k\bar a_l^*\Delta^{(1)}_{jkl;m}\\ 
  -2i\bar\eta\sin{q_m\over2} \sum_{j,k,l}r_l\left[2 \bar a_j^* \bar
  a_k \bar a_l - \bar a_j \bar a_k \bar a_l^*\right]\sin{q_j\over2}
  \sin{q_k\over2} \sin{q_l\over2} \Delta^{(2)}_{jkl;m} =0.   
\end{eqnarray}

This is the main result of the perturbative calculation. We have
managed to replace $N$ coupled differential equations (\ref{eq:eom})
for the oscillator coordinates $x_n(t)$ by $N$ coupled algebraic
equations (\ref{eq:ampeq}) for the time-independent mode amplitudes
$a_m$.  All that remains, in order to obtain the overall collective
response of the array as a function of the parameters of the original
EOM, is to solve these coupled algebraic equations.

Before doing so we should note the following general statements.
First, one can easily verify that for a single oscillator
($N=j=k=l=m=1$), the general equation (\ref{eq:ampeq}) reduces to the
single-oscillator equation (\ref{eq:amp1}), we derived in
section~\ref{one}.  Next, one can also see that the trivial solution,
$a_m=0$ for all $m$, always satisfies the equations, though, as we
have seen in the case of a single oscillator, it is not always a
stable solution. Finally, one can also verify that whenever for a
given $m$, $\Delta^{(1)}_{mmm;j} = \Delta^{(2)}_{mmm;j} = 0$ for all
$j\neq m$, then a single-mode solution exists with $a_m\neq 0$ and
$a_j=0$ for all $j\neq m$. These single-mode solutions have the
elliptical shape of the single-oscillator solution given in
Eq.~(\ref{eq:response1}), and satisfy the equation
\begin{equation}
  \label{eq:singlemode}
  {1\over{4\sin^4(q_m/2)}}\left({3\over4}\Delta^{(1)}_{mmm;m}|\bar
  a_m|^2 - \bar\Omega\right)^2 + \left(1+ \sin^2{q_m\over2}
  \Delta^{(2)}_{mmm;m}\bar\eta|\bar a_m|^2\right)^2 = \bar h^2, 
\end{equation}
where for each solution we have set $\omega_0=\omega_m$, so that
$\delta_m=0$ and $r_m=1$. Note that generically $\Delta^{(1)}_{mmm;m} =
\Delta^{(2)}_{mmm;m} = 3$, except when umklapp conditions are satisfied.

Additional solutions, involving more than a single mode, exist in
general but are hard to obtain analytically. We calculate these
multi-mode solutions below for the case of two and three oscillators
by finding the roots of the coupled algebraic equations numerically.
In Appendix~\ref{explicit} we show the explicit sets of coupled
mode-amplitude equations for these cases.

\begin{figure}
\includegraphics*[height=7.0in]{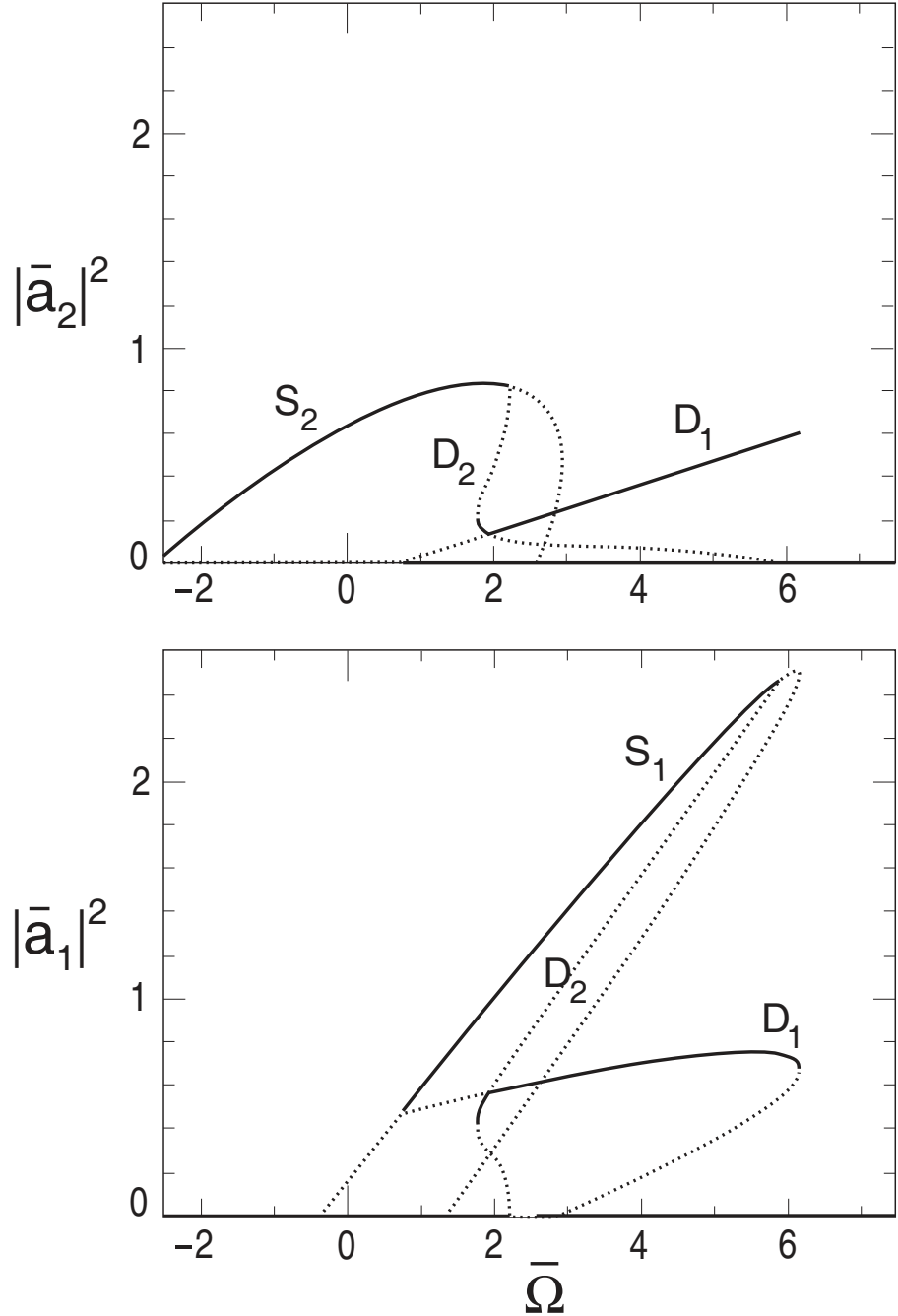}%
\caption{\label{2beams}
  Two oscillators: Response intensity of as a function of frequency
  $\bar\Omega$, for a particular choice of the equation parameters.
  The top graph shows $|\bar a_2|^2$, the bottom graph shows $|\bar
  a_1|^2$.  Solid curves indicate stable solutions and dashed curves
  indicate unstable solutions. The two elliptical single-mode solution
  branches [Equations (\ref{eq:s1}) and (\ref{eq:s2})] are labeled
  $S_1$ and $S_2$.  The two double-mode solution branches are labeled
  $D_1$ and $D_2$.}
\end{figure}

In Fig.~\ref{2beams} we show the solutions for the response intensity
of two oscillators as a function of frequency, for a particular choice
of the equation parameters. The top graph shows the square of the
amplitude of the antisymmetric mode $\bar a_2$, whereas the bottom
graph shows the square of the amplitude of the symmetric mode $\bar
a_1$. Solid curves indicate stable solutions and dashed curves
indicate unstable solutions. The two elliptical single-mode solution
branches, mentioned in the previous paragraph are easily spotted.
These branches are labeled by $S_1$ and $S_2$ (In
Appendix~\ref{explicit}, Equations (\ref{bothsingles}), we give the
analytical expressions for these two solution branches). In addition,
we find two double-mode solution branches, labeled $D_1$ and $D_2$,
involving the excitation of both modes simultaneously.  Note that the
two branches of double-mode solutions intersect at a point where they
switch their stability.

\begin{figure}
\includegraphics{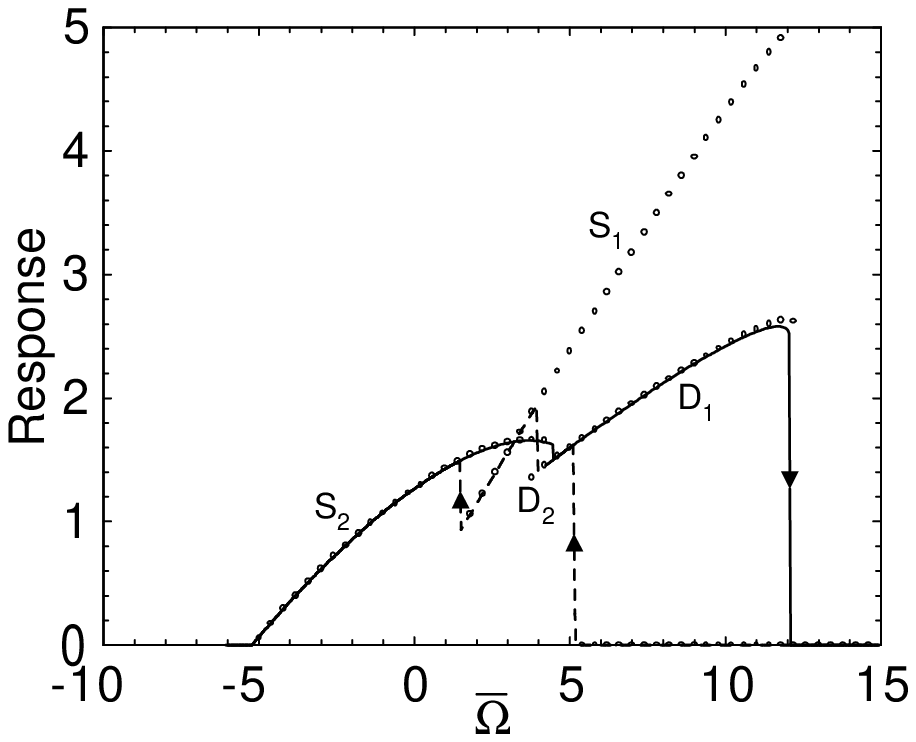}%
\caption{\label{2beamshyst}
  Hysteresis with two oscillators: Comparison of stable solutions,
  obtained analytically (small circles), with a numerical integration
  of the equations of motion (solid curve - frequency swept up;
  dashed curve - frequency swept down). Plotted is the averaged
  response intensity, defined in Eq.~(\ref{eq:intensity}). Branch
  labels correspond to those in Fig.~\ref{2beams}. }
\end{figure}

With two oscillators we obtain regions in frequency where three stable
solutions can exist. If all the stable solution branches are
accessible experimentally then the observed effects of hysteresis
might be more complex than in the simple case of a single oscillator.
This is demonstrated in Fig.~\ref{2beamshyst} where we compare our
analytical solutions with a numerical integration of the differential
equations of motion (\ref{eq:eom}) for two oscillators. The response
intensity, plotted here, is the time and space averages of the square
of the oscillator displacements
\begin{equation}
 \label{eq:intensity}
 I={1\over N}\sum_{n=1}^N \left\langle x_{n}^{2}\right\rangle,
\end{equation}
where the angular brackets denote time average, and here $N=2$. A
solid curve shows the response intensity for frequency swept upwards,
and a dashed curve shows the response intensity for frequency swept
downwards.  Small circles show the analytical response intensity,
using the fact that $I=3(|\bar a_1|^2 + |\bar a_2|^2)/2$, for the
stable regions of the four solution branches shown in
Fig.~\ref{2beams}.  With the analytical solution in the background,
one can easily understand all the discontinuous jumps, as well as the
hysteresis effects, that are obtained in the numerical solution of the
equations of motion. Note the the $S_1$ branch is missed in the
upwards frequency sweep and is only accessed by the system in the
downwards frequency sweep. One could trace the whole stable region of
the $S_1$ branch by changing the sweep direction after jumping onto
the branch, thereby climbing all the way up to the end of the $S_1$
branch and then falling onto the tip of the $D_1$ branch or to zero.

\begin{figure}
\includegraphics*[height=7.0in]{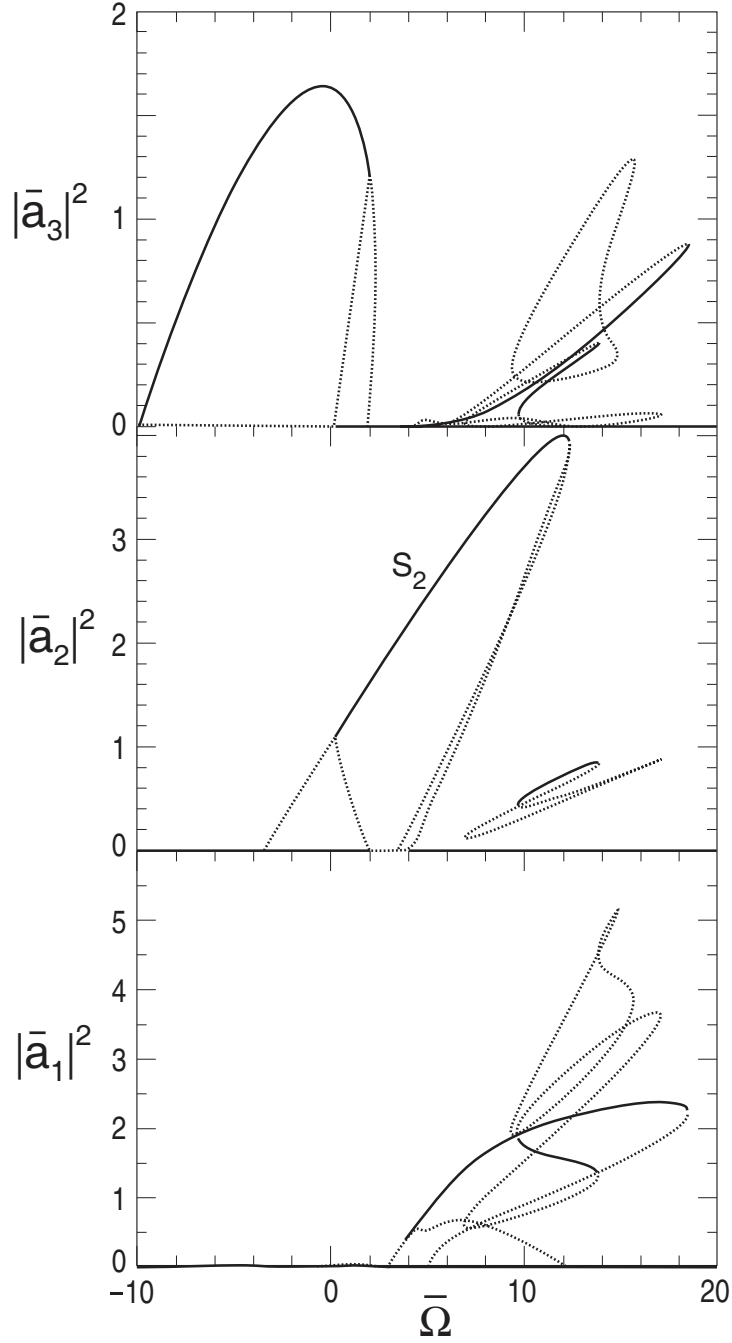}%
\caption{\label{3beams}
  Three oscillators: Response intensity of three oscillators as a
  function of frequency $\bar\Omega$, for a particular choice of the
  equation parameters. The graphs show the squares of the amplitudes
  of the three different modes. Solid curves indicate stable solutions
  and dashed curves indicate unstable ones. The only elliptical
  single-mode solution branch [Eq. (\ref{eq:3beams2})] is labeled by
  $S_2$.}
\end{figure}

In Fig.~\ref{3beams} we show the solutions for the response intensity
of three oscillators as a function of frequency, for a particular
choice of the equation parameters. The graphs show the squares of the
amplitudes of the three different modes. Solid curves indicate stable
solutions and dashed curves indicate unstable ones. For three
oscillators there is only one elliptical single-mode solution branch,
of the form of Eq.~(\ref{eq:singlemode}), whose exact analytical
expression is given in Eq. (\ref{eq:3beams2}). This branch is labeled
by $S_2$. In addition, we find a host of nontrivial multi-mode
solution branches, including one that is disconnected from all other
branches. We show these plots, not only to demonstrate that it is
possible to obtain such solutions exactly, but also to emphasize the
large number and nontrivial structure of the solution branches one
finds, even for such a small number of oscillators. This can only
serve as a hint for the multi-mode solutions one can expect to find when
the number of oscillators is large, as in the BR experiment.

\section{\label{many}
  Response of parametrically-driven nonlinear coupled 
  oscillators---Numerical integration of the equations}

\begin{figure}
\begin{center}
\includegraphics{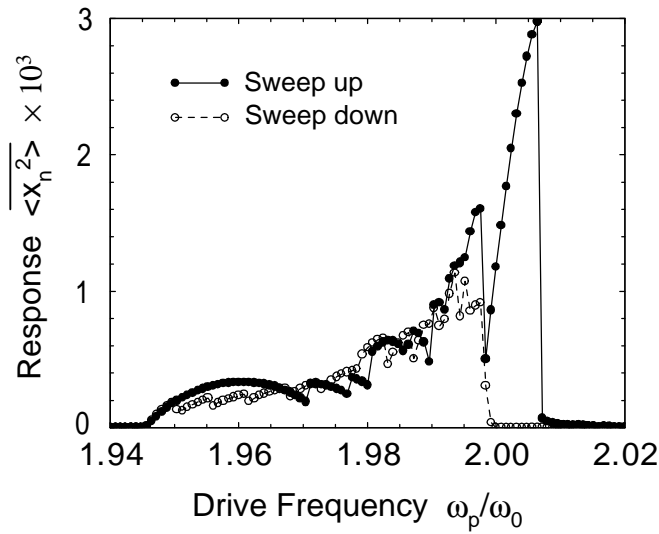}%
\caption{Response intensity as a function of the driving frequency
  $\omega_p$ (measured in units of the top band edge frequency
  $\omega_0$) for $N=67$ parametrically-driven oscillators (solid
  curve - frequency swept up; dashed curve - frequency swept down).
  The response intensity is defined as $\left\langle
    \overline{x_{n}^{2}}\right\rangle$ [Eq.~(\ref{eq:intensity})],
  with the bar denoting the average over the space index $n$, and the
  brackets the average over time. The
parameters used are $\Delta^{2}=0.02,h=0.16,\gamma=0.004,\eta=6$.}%
\label{67beams}%
\end{center}
\end{figure}
%EndExpansion

The equations of motion (\ref{eq:eom}) were integrated numerically for
an array of $67$ oscillators, as in the BR experiment. The results
for the response intensity (\ref{eq:intensity}) as a function of
parametric drive frequency $\omega_p$ (measured in units of the top
band edge frequency $\omega_0$) are shown in Figure \ref{67beams}.
These results must be considered illustrative only, since many of the
parameters of the experimental system are not known.  The parameters
used to construct the figure, $\Delta^{2}=0.02,h=0.016,\gamma
=0.004,\eta=6.0$, were chosen using the insights gained from the two
and three beam results. We should emphasize that the structure of the
response branches depends very strongly on the equation parameters.

A number of the important features of the experimental data are
reproduced by these calculations. We concentrate on the solid curve in
the figure, which is for frequency swept upwards, since this is the
protocol that was used in the experiment. In particular, the response
intensity shows features that span a range of frequencies that is
large compared with the mode spacing (which is about 0.0006 for the
parameter used). The lowest frequency feature, from about
$\omega_p/\omega_0=1.94$ to $\omega_p/\omega_0=1.97$, can be
identified as the response to the parametric drive of a single mode at
or very near the band edge at $\omega/\omega_0=0.98$, analogous to the
one mode response shown in Figure \ref{resp1omega}.  Furthermore, the
variation of the response with frequency shows abrupt jumps,
particularly on the high frequency side of the features as the
frequency is raised.  Finally, the response extends to frequencies
higher than the band edge for the linear modes, which would give a
response only up to $\omega_p/\omega_0=2.0$. All of these features are
understood now that we have seen the analytical solutions for small
numbers of oscillators. In particular, the wide features compared with
the mode spacing are explained by the simple fact the as the frequency
is swept upwards a particular solution branch is followed as long as
it remains stable.  In the meantime many other stable solutions, that
may be as close to each other as the mode spacing, are simply skipped
over.

Comparing the two traces in Figure \ref{67beams} shows that the
response for a downward frequency sweep is significantly different,
with a less dramatic variation of the response. In particular, note
that the downwards sweep was able to access additional stable solution
branches that were missed in the upwards sweep. There is also no
response above $\omega_p/\omega_0=2.0$ in this case. This is because
the zero displacement state is stable for $\omega_p/\omega_0>2.0$, and
the system will remain in this state as the frequency is lowered,
unless a large enough disturbance kicks it onto another of the
solution branches. The hysteresis on reversing the frequency sweep was
not looked at in the first experiments, and it would be interesting to
test this prediction in further experiments.

\section{\label{conc}
  conclusions}

We have calculated the response of nonlinear coupled oscillators to
parametric excitation. Our calculations agree qualitatively with the
experimental measurements of Buks and Roukes~\cite{buks2} and explain
the main features observed in the experiment. The abrupt drops in the
response as the frequency is swept upwards, the response continuing
beyond the upper edge of the frequency band, and the large size of the
response features compared with the mode spacing, are all
qualitatively explained.

Nevertheless, we propose that a more systematic study be conducted on
systems of coupled nonlinear resonators so that our theoretical
predictions could be tested more quantitatively. For example,
successive measurements on systems containing only one, two, and three
coupled resonators, that are made as identically as possible, could be
used to extract the nonlinear parameters of the resonators. These
could then be used to predict and explain the response of a large
resonator array more quantitatively.

Furthermore, we have demonstrated that as the number of oscillators is
increased the number of the solution branches for the response of the
system increases and the effects of hysteresis become more and more
complicated.  This suggests that the appropriate experimental protocol
for studying a system with many oscillators should be---in addition to
the standard up-sweep and down-sweep in frequency---to change the
direction of the sweep after every abrupt change in the response
intensity. Such a protocol may provide more information about the
response curve by accessing additional branches of the solution and
fully tracing them out.

\appendix

\section{\label{ap1}
  Parametric excitation of a single oscillator at its second
  instability tongue}

For a single nonlinear oscillator, like the one studied in
section~\ref{one}, which is parametrically-excited at its second
instability tongue, Eq.~(\ref{eq:preeom}) becomes
\begin{eqnarray}
  \label{eq:second}\nonumber
  \ddot x &+ &\bigl[\omega^2 - \Delta^2H\cos(\omega +
  \epsilon\Omega) t\bigr] x \\
  &+ &\epsilon\gamma \dot x + x^3 + \eta x^2 \dot x = 0,
\end{eqnarray}
where again $\omega=\sqrt{1-\Delta^2}$ is the resonance frequency of
the oscillator in the harmonic approximation, but the parametric
excitation is performed around $\omega$ and not around $2\omega$. In
this case the scaling of $\Delta^2H$ with respect to $\epsilon$ needs
to be redetermined. The technical reason for this is that if we naively
take $\Delta^2H=\epsilon h$, as before, then the parametric driving
term does not contribute to the order $\epsilon^{3/2}$ secular term
which we use to find the response, and the order $\epsilon^{1/2}$ term
in $x$ becomes identically zero.

The remedy for this situation is to let $\Delta^2H$ scale like
$\epsilon^p$, with $p<1$, so that there will be a non-secular
correction to $x$ at a lower order than $\epsilon^{3/2}$. The value of
$p$ will be chosen such that this correction will contribute to the
order $\epsilon^{3/2}$ secular term and will give us the required
response. The equation of motion (\ref{eq:second}) becomes
\begin{equation}
  \label{eq:secondh}
  \ddot x + \omega^2 x = {{h\epsilon^p}\over2}\left(e^{i(\omega t +
  \Omega T)} + c.c.\right)x - \epsilon\gamma\dot x - x^3 - \eta x^2\dot x,   
\end{equation}
and we try an expansion of the solution of the form
\begin{equation}
  \label{eq:ansatz2}
  x(t) = \epsilon^{1/2}\left(A(T) e^{i\omega t} + c.c.\right) 
       + \epsilon^{p+1/2}x_p(t) + \epsilon^{3/2}x_1(t) + \ldots
\end{equation}
Substituting this expansion into the equation of motion
(\ref{eq:secondh}) we obtain at order $\epsilon^{1/2}$ the linear
equation as usual, and at order $\epsilon^{p+1/2}$
\begin{equation}
  \label{eq:orderp}
  \ddot x_p + \omega^2 x_p = {h\over2}\left(Ae^{i(2\omega t +
  \Omega T)}+ A^*e^{i\Omega T} + c.c.\right).
\end{equation}
As expected, there is no secular term on the right-hand side, so we
can immediately solve for $x_p$,
\begin{equation}
  \label{eq:xp}
  x_p(t) = {h\over2}\left(-{A\over{3\omega^2}}e^{i(2\omega t +
  \Omega T)}+ {A^*\over{\omega^2}}e^{i\Omega T} + c.c.\right) + O(\epsilon).
\end{equation}
Substituting the solution for $x_p$ into the expansion
(\ref{eq:ansatz2}), and the expansion back into the equation of motion
(\ref{eq:secondh}), contributes an additional term from the parametric
driving which has the form
\begin{eqnarray}
  \label{eq:pterm}\nonumber
  \epsilon^{2p+1/2}{h^2\over4}\left(-{A\over{3\omega^2}}e^{i(2\omega t +
  \Omega T)}+ {A^*\over{\omega^2}}e^{i\Omega T} + c.c.\right)
  \left(e^{i(\omega t + \Omega T)} + c.c.\right)\\
  =\epsilon^{2p+1/2}{h^2\over{4\omega^2}}\left({2\over3}A +
  A^*e^{i2\Omega T}\right)e^{i\omega t} + c.c. + {\rm\ non\
  secular\ terms}. 
\end{eqnarray}
To contribute to the order $\epsilon^{3/2}$ secular term, we see that
we must set $p=1/2$. This gives us the required contribution to the
equation for the vanishing secular terms. All other terms remain as
they were in Eq.~(\ref{eq:secular1}), so that the new equation for
determining $A(T)$ becomes
\begin{equation}
  \label{eq:secular2}
  2i\omega {dA\over dT} - {h^2\over{4\omega^2}}\left({2\over3}A +
  A^*e^{i2\Omega T}\right)
  + i\omega\gamma A + 3|A|^2 A + i\omega\eta |A|^2 A = 0.
\end{equation}
Again, ignoring initial transients, and assuming that the nonlinear
terms in the equation are sufficient to saturate the growth of the
instability, we try a steady-state solution, this time of the form
\begin{equation}
  \label{eq:A2}
  A(T) = a e^{i\Omega T}.
\end{equation}
The solution to the equation of motion~(\ref{eq:single}) is therefore
\begin{equation}
  \label{eq:sol2}
  x(t) = \epsilon^{1/2}(a e^{i(\omega+\epsilon\Omega) t} + c.c.) 
       + O(\epsilon),
\end{equation}
where the correction $x_{1/2}$ of order $\epsilon$ is given in
Eq.~(\ref{eq:xp}) and, as before, we are not interested in the
correction $x_1(t)$ of order $\epsilon^{3/2}$, but rather in the fixed
amplitude $a$ of the lowest order term. We substitute the steady-state
solution~(\ref{eq:A2}) into the equation~(\ref{eq:secular2}) of the
secular terms and obtain
\begin{equation}
  \label{eq:amp2}
  \left[\left(3|a|^2 - 2\omega\Omega -
  {2\over3}{{h^2}\over{4\omega^2}}\right) +  
  i\omega\left(\gamma + \eta |a|^2\right)\right]a =
  {{h^2}\over{4\omega^2}} a^*. 
\end{equation}

We divide both sides of the last equation by $\gamma\omega$ and define
the rescaled variables: $\bar a=a/\sqrt{\gamma\omega}$,
$\bar\Omega=\Omega/\gamma$, $\bar\eta=\omega\eta$, and $\bar
h=h/2\sqrt{\gamma\omega^3}$, in terms of which we obtain after taking the
magnitude squared of both sides, in addition to the trivial solution
$a=0$, the non-trivial response
\begin{equation}
  \label{eq:response2}
  \left({3|\bar a|^2} - 2\bar\Omega - {2\over3}\bar h^2\right)^2 
  + \left(1 + \bar\eta|\bar a|^2\right)^2 = \bar h^4.
\end{equation}

\begin{figure}
\includegraphics{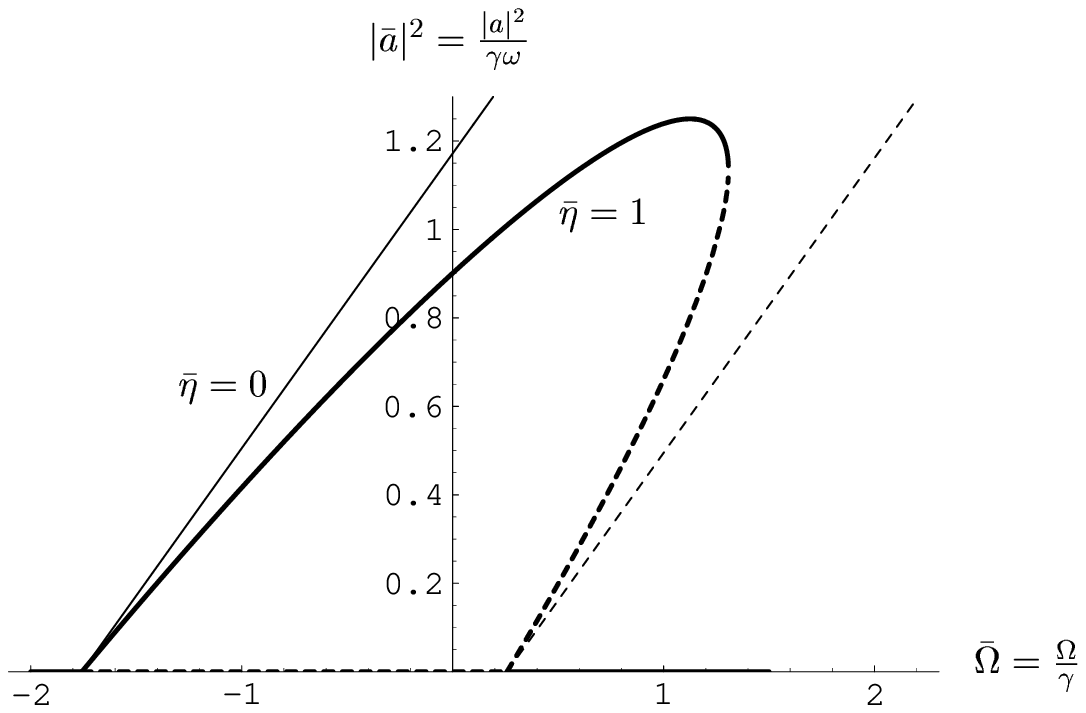}%
\caption{\label{tongue2}
  Response intensity $|\bar a|^2$ as a function of the frequency
  $\bar\Omega$, for fixed amplitude $\bar h=1.5$ in the second
  instability tongue.  Solid curves are stable solutions; dashed
  curves are unstable solutions. Thin curves show the response without
  non-linear damping ($\bar\eta=0$). Thick curves show the response
  for finite nonlinear damping ($\bar\eta=1$).}
\end{figure}

Fig.~\ref{tongue2} shows the response intensity $|\bar a|^2$ as a
function of the frequency $\bar\Omega$, for fixed amplitude $\bar
h=1.5$ in the second instability tongue. The solution looks very
similar to the response shown in Fig.~\ref{resp1omega} for the first
instability tongue, though we should point out two important
differences. The first is that the orientation of the ellipse,
indicated by the slope of the curves for $\bar\eta=0$, is different.
The slope here is $2/3$ whereas for the first instability tongue the
slope is $1/3$. The second is the change in the definition of $\bar
h$. The lowest amplitude required for having an instability, is again
on resonance ($\bar\Omega=0$), and its value is again $\bar h=1$, but
now this implies that $h=2\sqrt{\gamma\omega^3}$, or that $h$ it
scales as $\sqrt\gamma$. This is consistent with the well known result
(see, for example, Ref.~\onlinecite{landau}) that the minimal amplitude for
the instability of the $n^{th}$ tongue scales as $\gamma^{1/n}$.

\section{\label{explicit}
 Explicit equations for two and three coupled oscillators}

\subsection{Two coupled nonlinear oscillators}

For two coupled oscillators ($N=2$) we have
\begin{eqnarray}
  \label{eq:2params}
  &&q_1={\pi\over3},\qquad\qquad q_2={{2\pi}\over3}, \\
  &&\omega_1^2 = 1-{1\over2}\Delta^2,\qquad \omega_2^2 = 1-{3\over2}\Delta^2,
\end{eqnarray}
and we choose the reference frequency $\omega_0$ to be $\omega_2$, so
that $\delta_2=\Omega_2=0$, $r_2=1$,
$\delta_1=2(\omega_1-\omega_2)/\gamma\epsilon\equiv\delta>0$, and
$r_1=\omega_1/\omega_2\equiv r$. For $\Delta\ll1$, $\delta\simeq
\Delta^2/\epsilon\gamma$ and $r\simeq 1+\Delta^2/2$. The first mode is
the symmetric one with $x_1(t)=x_2(t)$ and the second mode is
antisymmetric with $x_1(t)=-x_2(t)$. The equations
(\ref{eq:ampscaled}) for the rescaled complex amplitudes $\bar a_1$
and $\bar a_2$ are
\begin{subequations}
\begin{eqnarray}
  \label{eq:2beams}\nonumber
  &&(\bar\Omega-\delta) r \bar a_1 -
  i {r\over2} \bar a_1 + {\bar h\over2} \bar a_1^*
  -{9\over4}\left(|\bar a_1|^2\bar a_1 + 2|\bar a_2|^2\bar a_1 +
  \bar a_2^2\bar a_1^*\right)\\ 
  &&\quad-{3\over8}i\bar\eta\left[r|\bar a_1|^2\bar a_1 + 2r|\bar
  a_2|^2\bar a_1 + (2-r)\bar a_2^2\bar a_1^*\right] =0,\\  \nonumber
  &&\bar\Omega \bar a_2 - i {3\over2} \bar a_2 +
  {3\over2}\bar h \bar a_2^* -
  {9\over4}\left(|\bar a_2|^2\bar a_2 + 2|\bar a_1|^2\bar a_2 +
  \bar a_1^2\bar a_2^*\right)\\
  &&\quad-{3\over8}i\bar\eta\left[9|\bar a_2|^2\bar a_2 + 2|\bar
  a_1|^2\bar a_2 + (2r-1)\bar a_1^2\bar
  a_2^*\right] =0. 
\end{eqnarray}
\end{subequations}

The two single-mode solution branches, having the general form of
Eq.~(\ref{eq:singlemode}) and labeled $S_1$ and $S_2$ in
Fig.~\ref{2beams}, are easily obtained by setting $\bar a_2$ or $\bar
a_1$ to zero in the coupled equations above, respectively. This yields
the analytical forms of these solutions which are 
\begin{subequations}\label{bothsingles}
\begin{eqnarray}
  \label{eq:s1}
  &S_1:\quad&\left({{9\over2}|\bar a_1|^2} - 2r(\bar\Omega-\delta)\right)^2 
  + r^2\left(1 + {3\over4}\bar\eta|\bar a_1|^2\right)^2 = \bar
  h^2,\\\label{eq:s2} 
  &S_2:\quad&\left({{3\over2}|\bar a_2|^2} - {2\over3}\bar\Omega\right)^2 
  + \left(1 + {9\over4}\bar\eta|\bar a_2|^2\right)^2 = \bar h^2.
\end{eqnarray}
\end{subequations}

\subsection{Three coupled nonlinear oscillators}

For three coupled oscillators ($N=3$) we have
\begin{eqnarray}
  \label{eq:3params}
  &&q_1={\pi\over4},\qquad\qquad q_2={\pi\over2},\qquad\qquad
  q_3={{3\pi}\over4} \\ 
  &&\omega_1^2 = 1-\Delta^2 + \Delta^2/\sqrt2,
    \quad \omega_2^2 = 1-\Delta^2, 
    \quad \omega_3^2 = 1-\Delta^2 - \Delta^2/\sqrt2,
\end{eqnarray}
and we choose the reference frequency $\omega_0$ to be $\omega_2$, so
that $\delta_2=0$, $r_2=1$, and
$\delta_1=2(\omega_1-\omega_2)/\gamma\epsilon=-\delta_3\equiv\delta>0$.
For $\Delta\ll1$, $\delta\simeq \Delta^2/\sqrt2\epsilon\gamma$
and $r_{1,3}\simeq 1\pm\Delta^2/2\sqrt2$. The equations (\ref{eq:ampscaled}) for the
rescaled complex amplitudes $\bar a_1$, $\bar a_2$, and $\bar a_3$ are
\begin{subequations}
\begin{eqnarray}
  \label{eq:3beams}\nonumber
  &&(\bar\Omega-\delta) r_1 \bar a_1 -
  i {{2-\sqrt2}\over2}r_1 \bar a_1 + {{2-\sqrt2}\over2}\bar h \bar
  a_1^*\\ \nonumber
  &&-{3\over4}\left(3|\bar a_1|^2\bar a_1  - |\bar a_3|^2\bar a_3 
  + 4|\bar a_2|^2\bar a_3 + 2\bar a_2^2\bar a_3^* 
  + 4|\bar a_2|^2\bar a_1 + 2\bar a_2^2\bar a_1^*
  + 6|\bar a_3|^2\bar a_1 + 3\bar a_3^2\bar a_1^*
  - 2|\bar a_1|^2\bar a_3 - \bar a_1^2\bar a_3^*
  \right)\\ \nonumber
  &&\quad-i\bar\eta\left[{{3-2\sqrt2}\over4}3r_1|\bar a_1|^2\bar a_1
  - {{\sqrt2+1}\over4}r_3|\bar a_3|^2\bar a_3
  +{{2-\sqrt2}\over2}\left(2r_1|\bar a_2|^2\bar
  a_1 + (2-r_1)\bar a_2^2\bar a_1^*\right)\right.\\ 
  &&\qquad\left.+{{\sqrt2-1}\over4}\left(2r_3|\bar a_1|^2\bar a_3 
  + (2r_1-r_3)\bar a_1^2\bar a_3^*\right)
  +{1\over4}\left(2r_1|\bar a_3|^2\bar a_1 +
  (2r_3-r_1)\bar a_3^2\bar a_1^*\right)
  \right]=0,\\  \nonumber
  &&\bar\Omega \bar a_2 - i\bar a_2 + \bar h \bar a_2^* -
  {3\over2}\left(2|\bar a_2|^2\bar a_2 + 2|\bar a_1|^2\bar a_2 +
  \bar a_1^2\bar a_2^* + 2|\bar a_3|^2\bar a_2 +
  \bar a_3^2\bar a_2^* + \bar a_1^*\bar a_2\bar a_3 + \bar a_1\bar
  a_2^*\bar a_3 + \bar a_1\bar a_2\bar a_3^* \right)\\ \nonumber
  &&\quad-i\bar\eta\left[|\bar a_2|^2\bar a_2 
  +{{2-\sqrt2}\over2}\left(2|\bar a_1|^2\bar a_2 + (2r_1-1)\bar
  a_1^2\bar a_2^*\right)\right.\\ 
  &&\qquad\left.+ {{2+\sqrt2}\over2}\left(2|\bar a_3|^2\bar a_2 
  + (2r_3-1)\bar a_3^2\bar a_2^*\right)\right] =0,\\ \nonumber
  &&(\bar\Omega+\delta) r_3 \bar a_3 -
  i {{2+\sqrt2}\over2}r_3 \bar a_3 + {{2+\sqrt2}\over2}\bar h \bar
  a_3^*\\ \nonumber
  &&-{3\over4}\left(3|\bar a_3|^2\bar a_3  - |\bar a_1|^2\bar a_1 
  + 4|\bar a_2|^2\bar a_3 + 2\bar a_2^2\bar a_3^* 
  + 4|\bar a_2|^2\bar a_1 + 2\bar a_2^2\bar a_1^*
  + 6|\bar a_1|^2\bar a_3 + 3\bar a_1^2\bar a_3^*
  - 2|\bar a_3|^2\bar a_1 - \bar a_3^2\bar a_1^*
  \right)\\ \nonumber
  &&\quad-i\bar\eta\left[{{3+2\sqrt2}\over4}3r_3|\bar a_3|^2\bar a_3
  + {{\sqrt2-1}\over4}r_1|\bar a_1|^2\bar a_1
  +{{2+\sqrt2}\over2}\left(2r_3|\bar a_2|^2\bar
  a_3 + (2-r_3)\bar a_2^2\bar a_3^*\right)\right.\\ 
  &&\qquad\left.-{{\sqrt2+1}\over4}\left(2r_1|\bar a_3|^2\bar a_1 
  + (2r_3-r_1)\bar a_3^2\bar a_1^*\right)
  +{1\over4}\left(2r_3|\bar a_1|^2\bar a_3 +
  (2r_1-r_3)\bar a_1^2\bar a_3^*\right)\right]=0.
\end{eqnarray}
\end{subequations}

Only one single-mode solution of the form of Eq.~(\ref{eq:singlemode})
exists in the case of three oscillators, and involves the second
mode. It is obtained by setting $\bar a_1=\bar a_3=0$ in the coupled
equations above. The analytical expression for this solution is
\begin{equation}
  \label{eq:3beams2}
  S_2:\quad \left({3|\bar a_2|^2} - \bar\Omega\right)^2 
  + \left(1 + \bar\eta|\bar a_2|^2\right)^2 = \bar h^2.
\end{equation}

\begin{acknowledgments}
  This work was supported by Grant No.~1999458 from the U.S.-Israel
  Binational Science Foundation (BSF), and NSF Grant No.~DMR-9873573.
  We thank Eyal Buks, Michael Roukes, and Inna Kozinsky for many
  useful discussions and for sharing their experimental results prior
  to publication.
\end{acknowledgments}

\bibliography{beams}

\end{document}